\documentclass[12pt]{article}
\pdfoutput=1
\usepackage{amssymb,amsmath}
\usepackage{graphicx}
\usepackage{color}
\usepackage[colorlinks=true
,urlcolor=blue
,citecolor=blue
,linkcolor=blue
,pagecolor=blue
,linktocpage=true
,pdfproducer=medialab
]{hyperref}
\usepackage[a4paper]{geometry}

\usepackage{cite}
\usepackage{placeins}
\usepackage{slashed}
\usepackage{mathtools}
\usepackage[FIGTOPCAP]{subfigure}

\makeatletter \renewcommand{\@dotsep}{10000} \makeatother

\def\tst{\tilde t}

\mathchardef\mhyphen="2D
\def\tbta{t\mhyphen b\mhyphen\tau}
\def\btau{b\mhyphen\tau}
\def\mgut{M_{{\rm GUT}}}
\def\psg{SU(4)_{c}\times SU(2)_{L}\times SU(2)_{R}}
\def\422{$4\mhyphen2\mhyphen2$}

\newcommand{\beq}{\begin{equation}}
\newcommand{\eeq}{\end{equation}}
\newcommand{\bea}{\begin{eqnarray}}
\newcommand{\eea}{\end{eqnarray}}

\newcommand\prd[3]{{\it Phys.\ Rev.\ }{\bf D #1} (#2) #3}
\newcommand\prl[3]{{\it Phys.\ Rev.\ Lett.\ }{\bf #1} (#2) #3}

\newcommand\jhep[3]{{\it J. High Energy Phys.\ }{\bf #1} (#2) #3}

\usepackage[nodisplayskipstretch]{setspace}

\begin{document}

\begin{titlepage}
\pagestyle{empty}

\vspace*{0.2in}
\begin{center}
{\Large \bf  Testing Yukawa Unification at LHC Run-3 and HL-LHC}\\
\vspace{1cm}
{\bf  Mario E. G\'omez$^a$\footnote{Email: mario.gomez@dfa.uhu.es} 
Qaisar Shafi$^b$\footnote{Email: shafi@bartol.udel.edu} 
 and
Cem Salih $\ddot{\rm U}$n$^a$\footnote{E-mail: cemsalihun@uludag.edu.tr}}
\vspace{0.5cm}

{\it $^a$Departamento de Ciencias Integradas y Centro de Estudios Avanzados en F\'{i}sica Matem\'aticas y Computación, Campus del Carmen, Universidad de Huelva, Huelva 21071, Spain.} \\ 

{\it $^b$ Bartol Research Institute, Department of Physics and Astronomy, University of Delaware, Newark, DE 19716, USA} \\

{\it $^c$Department of Physics, Bursa Uluda\~{g} University, TR16059 Bursa, Turkey}

\end{center}

\vspace{0.5cm}
\begin{abstract}
\noindent

We explore $\tbta$ Yukawa unification (YU) in a
supersymmetric $SU(4)_c \times SU(2)_L \times SU(2)_R $ {model} without imposing a discrete left-right (L-R) {symmetry}. A number of interesting solutions that are compatible with $\tbta$ YU, LSP neutralino dark matter (DM), and LHC and other experimental constraints are identified. In particular, they include gluino-neutralino and stau-neutralino co-annihilation scenarios, where the NLSP gluino mass can range from {1-3} TeV. Higgsino-like dark matter solutions are also identified for which gluino masses can approach 5 TeV or so. This scenario will be tested at LHC Run-3 and its future upgrades.

\end{abstract}

\end{titlepage}

\setcounter{footnote}{0}

\section{Introduction}
\label{ch:introduction}

Third family $\tbta$ Yukawa Unification (YU) \cite{big-422} arises naturally in the simplest versions of 
supersymmetric SO(10) and $\psg$ (\422) models and it has attracted a fair amount of attention in recent years \cite{bigger-422}. Most work on the implications of YU have assumed the presence of a discrete Left-Right (LR) symmetry ({more} precisely C-parity, also known as D-parity), which restricts the number of Soft Supersymmetry Breaking (SSB) parameters. It was also shown that $\tbta$ YU allows {the} gluino NLSP solutions in \422 \cite{Gogoladze:2009ug} {models with LR symmetry}. {In this case} the LSP neutralino {can provide the desired} dark matter (DM) {abundance, but} the gluino {turns out to be not much heavier than a} TeV {or so}. {Switching from} $\tbta$ YU to $\btau$ YU also allows stop NLSP solutions with $m_{\tilde{t}_{1}} \lesssim 1$ TeV \cite{Raza:2014upa}. 

The spontaneous breaking of $SO(10)$ to its maximal subgroup \422 can be accomplished either with a Higgs 54-plet or 210-plet. The breaking with 54-plet leaves the LR symmetry unbroken \cite{Kibble:1982dd, Lazarides:1985my,Lazarides:2019xai}. However, using the 210-plet yields \422 symmetry but the C-parity in this case is absent \cite{Chang:1983fu}. The spontaneous breaking of LR symmetry also avoids a potential domain wall problem \cite{Lazarides:1985my}. Recent works \cite{Gomez:2018zzw} have discussed the sparticle spectroscopy, DM implications and muon $g-2$ in \422 with broken LR symmetry in the softly broken scalar sector without imposing the $\tbta$ YU condition. 

In the case of broken LR symmetry the universality between the $SU(2)_{L}$ and $SU(2)_{R}$ gauginos does not hold, i.e. $M_{2L}\neq M_{2R}$. Besides, the symmetrical structure of \422 also allows non-universality among the other gauginos as

\begin{equation}
M_{1}= \dfrac{3}{5}M_{2R}+\dfrac{2}{5}M_{3}~.
\label{eq:gauginomasses}
\end{equation}

{Despite} non-universal gaugino masses, the gauge coupling unification can be maintained if \422 breaks to the MSSM gauge group {at the} grand unification scale ($\mgut$). We should also note the fact that the presence of Higgs 210-plet, in general, breaks YU \cite{Babu:1992ia}. However, $\tbta$ YU can be {largely} preserved {if} the third family matter fields {acquire masses from the $(1,2,2)$ components of the effective MSSM Higgs doublets} \cite{Ajaib:2013zha}.

In this paper we explore the low energy consequences of imposing  t-b-$\tau$ YU in a supersymmetric \422 model without assuming LR symmetry in the softly broken scalar and gaugino sectors. We employ a variety of constraints from collider physics, rare $B-$meson decays and DM searches, and we require that the LSP neutralino saturates the dark matter limits set by the Planck satellite {experiment}. The rest of the paper is organized as follows. We briefly describe {in Section \ref{sec:scan}} the scanning procedure and the experimental constraints. Section \ref{sec:tbtau} discusses the  low energy implications {if} $\tbta$ YU is imposed at $\mgut$ {and} present {some} benchmark points to exemplify our findings. {In Section \ref{sec:conc} we summarize {our} conclusions}.

\section{Scanning Procedure and Experimental Constraints}
\label{sec:scan}

We employ the ISAJET~7.84 package~\cite{ISAJET} 
 to perform random scans over the parameter space 
 given below. 
In this package, the weak scale values of gauge and third 
 generation Yukawa couplings are evolved to 
 $M_{\rm GUT}$ via the MSSM renormalization group equations (RGEs)
 in the $\overline{DR}$ regularization scheme.
We do not strictly enforce the unification condition
 $g_3=g_1=g_2$ at $M_{\rm GUT}$, since a few percent deviation
 from unification can be assigned to unknown GUT-scale threshold
 corrections~\cite{Hisano:1992jj}.
With the boundary conditions given at $M_{\rm GUT}$, 
 all the SSB parameters, along with the gauge and Yukawa couplings, 
 are evolved back to the weak scale $M_{\rm Z}$.

In evaluating Yukawa couplings the SUSY threshold 
 corrections~\cite{Pierce:1996zz} are taken into account 
 at the common scale $M_{\rm SUSY}= \sqrt{m_{\tst_L}m_{\tst_R}}$. 
The entire parameter set is iteratively run between 
 $M_{\rm Z}$ and $M_{\rm GUT}$ using the full 2-loop RGEs
 until a stable solution is obtained.
To better account for leading-log corrections, one-loop step-beta
 functions are adopted for gauge and Yukawa couplings, and
 the SSB parameters $m_i$ are extracted from RGEs at appropriate scales
 $m_i=m_i(m_i)$.
The RGE-improved 1-loop effective potential is minimized
 at an optimized scale $M_{\rm SUSY}$, which effectively
 accounts for the leading 2-loop corrections.
Full 1-loop radiative corrections are incorporated
 for all sparticle masses.

We have scanned the parameter space of \422 with broken LR symmetry in both the scalar and gaugino {sectors}. The fundamental parameters in this framework and their ranges {are} as follows:

\begin{equation}
\begin{array}{lll}
0.1 \leq & m_{L} & \leq 10 ~{\rm TeV} \\
0.05 \leq & M_{2L} & \leq 5 ~{\rm TeV} \\
-3 \leq & M_{3} & \leq 5 ~{\rm TeV} \\
-3 \leq & A_{0}/m_{L} & \leq 3 \\
2 \leq & \tan\beta & \leq 65 \\
-3 \leq & {\rm x}_{{\rm LR}} & \leq 3 \\
-3 \leq & {\rm y}_{{\rm LR}} & \leq 3 \\
0 \leq & {\rm x}_{{\rm d}} & \leq 3 \\
-1 \leq & {\rm x}_{{\rm u}} & \leq 2~. \\
\end{array}
\label{paramSpacePSLR2}
\end{equation}
{Here} $m_{L}$ is the universal SSB mass term for the left-handed SUSY scalars, while $M_{2}$ and $M_{3}$ are the SSB gaugino mass terms. $A_{0}$ {denotes} the SSB trilinear scalar interaction term, and $\tan\beta$ is the ratio of the vacuum expectation values of the MSSM Higgs doublets such that $\tan\beta \equiv v_{u}/v_{d}$. ${\rm x}_{{\rm LR}}$ measures the LR breaking in the scalar sector {with} $m_{2R}={\rm x}_{{\rm LR}}m_{L}$, where $m_{2R}$ is the SSB mass term for the right-handed SUSY scalars. Similarly ${\rm y}_{{LR}}$ parametrizes the LR breaking in the gaugino sector as $M_{2R}={\rm y}_{{\rm LR}}M_{2L}$. We also employ non-universal SSB masses for the MSSM Higgs fields by setting $m_{H_{d}} = x_{d}m_{L}$ and $m_{H_{u}}=x_{u}m_{L}$.

In scanning the parameter space, we employ the Metropolis-Hastings
 algorithm as described in \cite{Belanger:2009ti}. 
The data points collected all satisfy the requirement of REWSB, 
 with the neutralino in each case being the LSP. 
After collecting the data, we impose the mass bounds on 
 all the particles \cite{Olive,Aad:2012tfa,Vami:2019slp} and 
 use the IsaTools package~\cite{bsg, bmm} and Ref.~\cite{mamoudi}
 to implement {the phenomenological} constraints from the rare $B-$meson decays \cite{Amhis:2012bh,Asner:2010qj} {and DM} observations \cite{Akrami:2018vks}. The {following experimental} constraints along with their uncertainties {are} employed in our analyses:

\begin{equation}
\setstretch{1.8}
\begin{array}{l}
m_h  = 123-127~{\rm GeV}
\\
m_{\tilde{g}} \geq 2.1~{\rm TeV}~(\geq 0.8~{\rm TeV~if~gluino~is~NLSP})
\\
0.8\times 10^{-9} \leq{\rm BR}(B_s \rightarrow \mu^+ \mu^-)
  \leq 6.2 \times10^{-9} \;(2\sigma)
\\
2.99 \times 10^{-4} \leq
  {\rm BR}(B \rightarrow X_{s} \gamma)
  \leq 3.87 \times 10^{-4} \; (2\sigma)
\\
0.15 \leq \dfrac{
 {\rm BR}(B_u\rightarrow\tau \nu_{\tau})_{\rm MSSM}}
 {{\rm BR}(B_u\rightarrow \tau \nu_{\tau})_{\rm SM}}
        \leq 2.41 \; (3\sigma) \\
   0.114 \leq \Omega_{{\rm CDM}}h^{2} \leq 0.126~(5\sigma)~.
\label{constraints}
\end{array}
\end{equation}

{In addition to these constraints, we quantify $\tbta$ YU with the parameter $R_{tb\tau}$ as}

\begin{equation}
R_{tb\tau}\equiv \dfrac{{\rm Max}(y_{t},y_{b},y_{\tau})}{{\rm Min}(y_{t},y_{b},y_{\tau})}
\end{equation}
{where $R_{tb\tau}=1$ means perfect $\tbta$ YU. However, considering various uncertainties we consider solutions to be compatible with $\tbta$ YU for $R_{tb\tau}\leq 1.1$.}
	
%

After the mass spectrum and DM implications are calculated, it is interesting to confront the predictions from the  models consistent with the constraints given in Eq.(\ref{constraints}) as well as $\tbta$ YU with the LHC bounds and {future} prospects. {To this end}, we {employ} the tools provided by  the Smodels-v1.2.2.\cite{Smodels,Kraml:2013mwa}. This package decomposes {the} theoretical models into the Simplified Model Spectra (SMS) which are compared with the data provided by the ATLAS and CMS collaborations \cite{SMS_cms,SMS_atlas}.  For each model, we use SUSY-HIT \cite{Djouadi:2006bz} to compute the decay ratios of the SUSY particles and  PYTHIA  \cite{Sjostrand:2014zea} to produce the corresponding cross sections. 

\section{$\tbta$ YU and DM Implications}
\label{sec:tbtau}

\begin{figure}[ht!]
\centering
\subfigure{\includegraphics[scale=1]{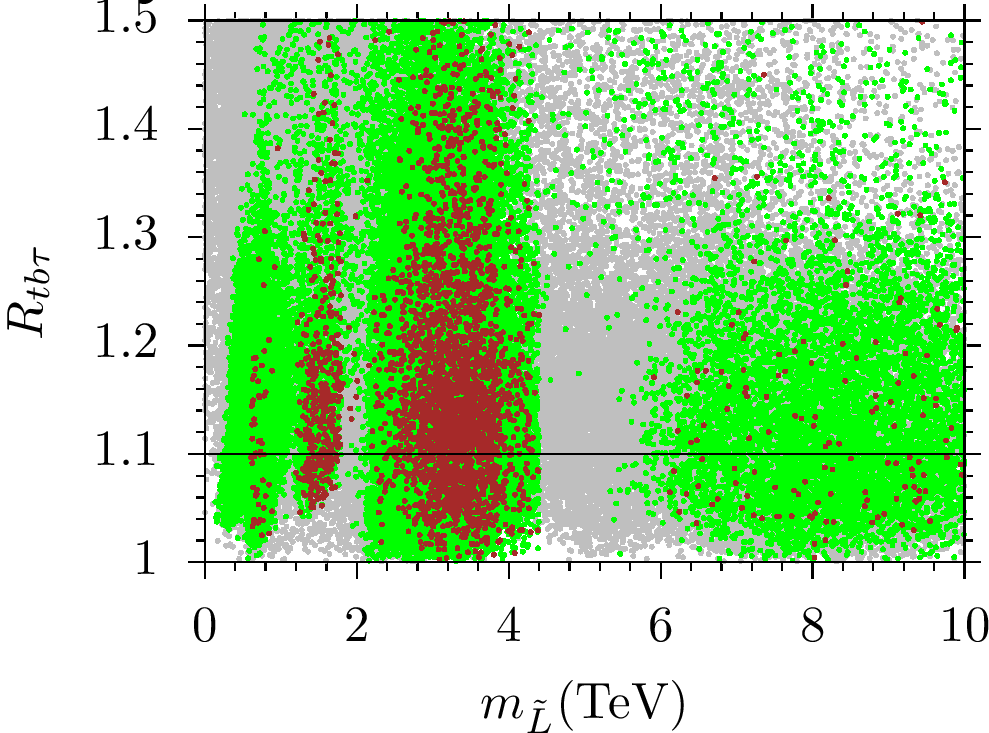}}
\subfigure{\includegraphics[scale=1]{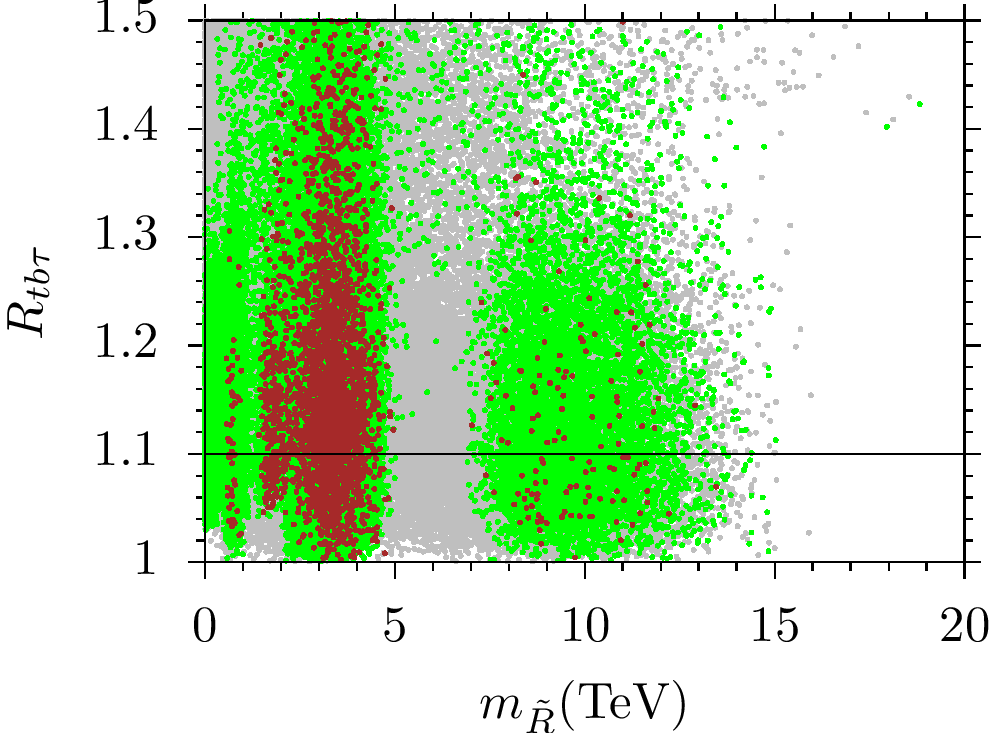}}
\subfigure{\includegraphics[scale=1]{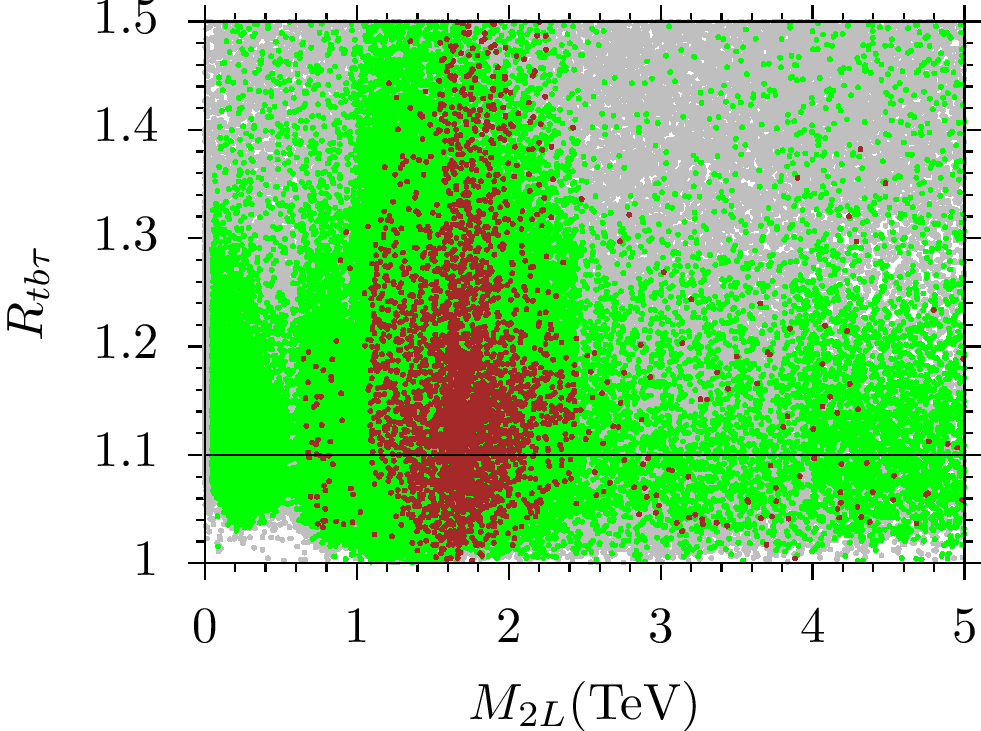}}
\subfigure{\includegraphics[scale=1]{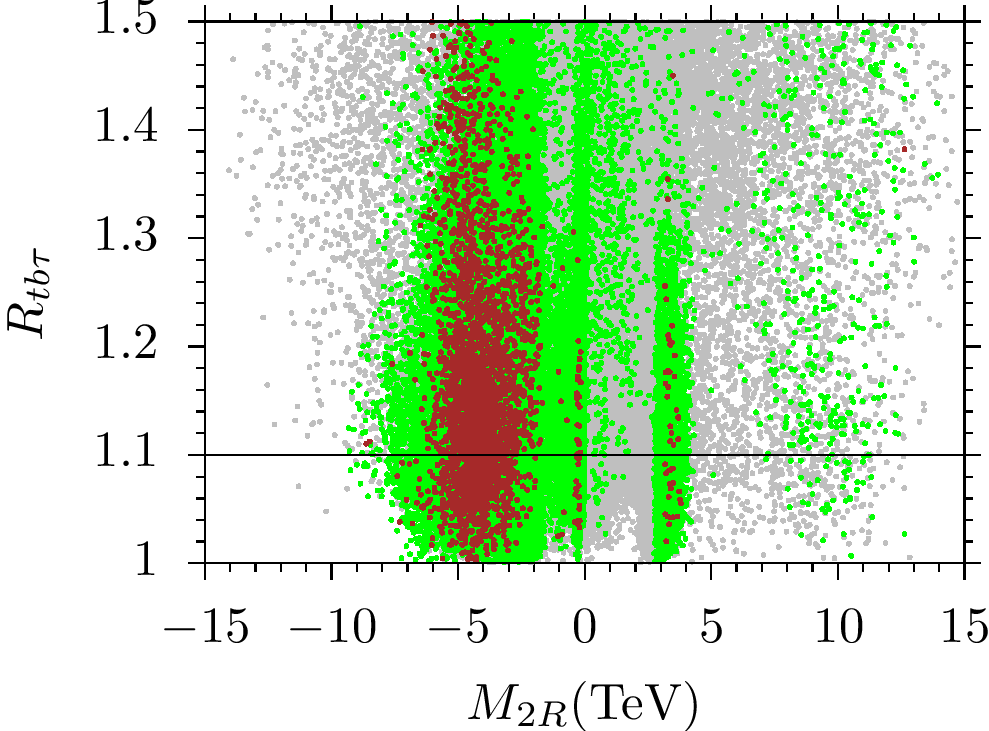}}
\caption{Plots in the $R_{tb\tau}-m_{\tilde{L}}$, $R_{tb\tau}-m_{\tilde{R}}$, $R_{tb\tau}-M_{2L}$ and $R_{tb\tau}-M_{2R}$ planes. All points are comparible with the REWSB and LSP neutralino conditions. Green points satisfy the mass bounds and constraints from rare $B-$meson decays. Brown points form a subset of green and they yield relic abundance of LSP neutralino consistent with the Planck measurements within $5\sigma$. {The regions below the horizontal lines correspond to $R_{tb\tau}=1.1$ are considered as to be compatible with the $\tbta$ Yukawa unification.}}
\label{fig1}
\end{figure}

\begin{figure}[ht!]
\centering
\subfigure{\includegraphics[scale=1]{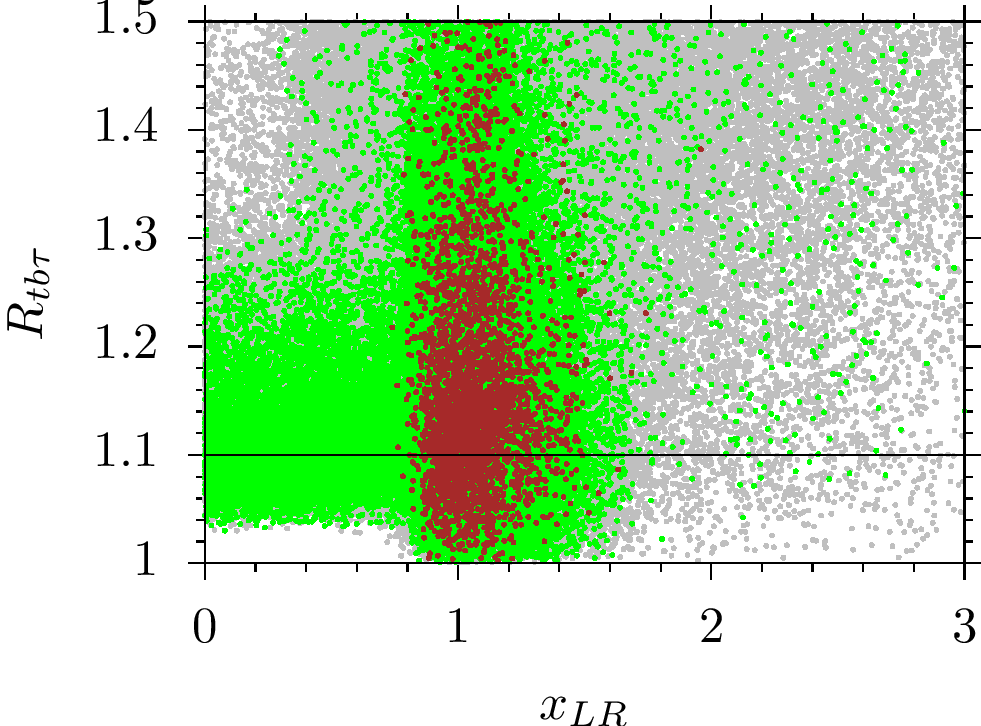}}
\subfigure{\includegraphics[scale=1]{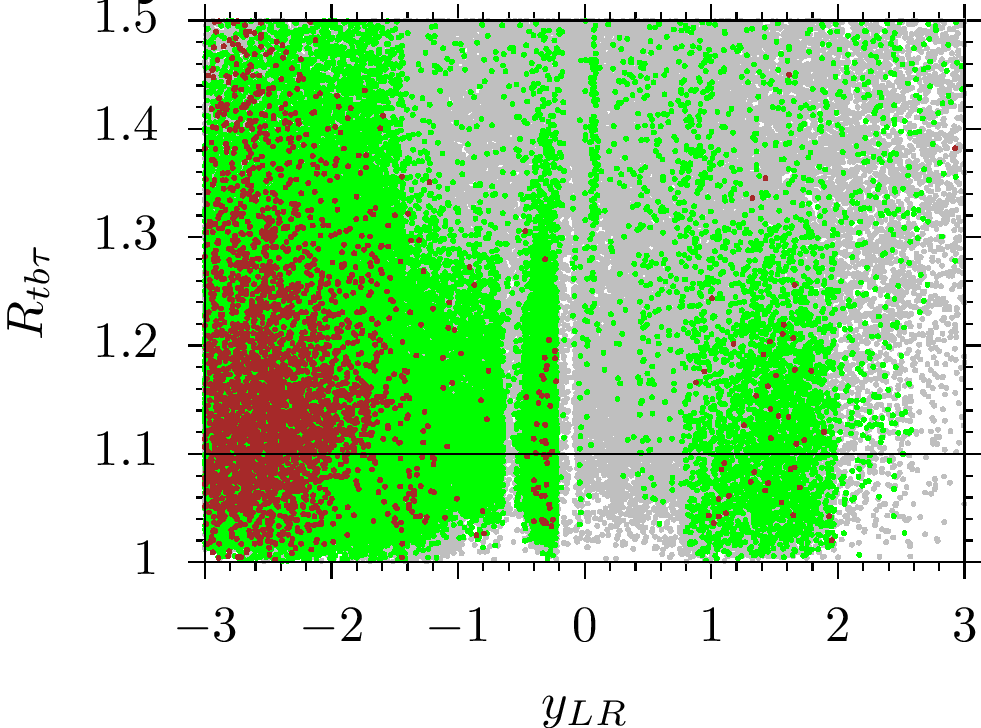}}
\subfigure{\includegraphics[scale=1]{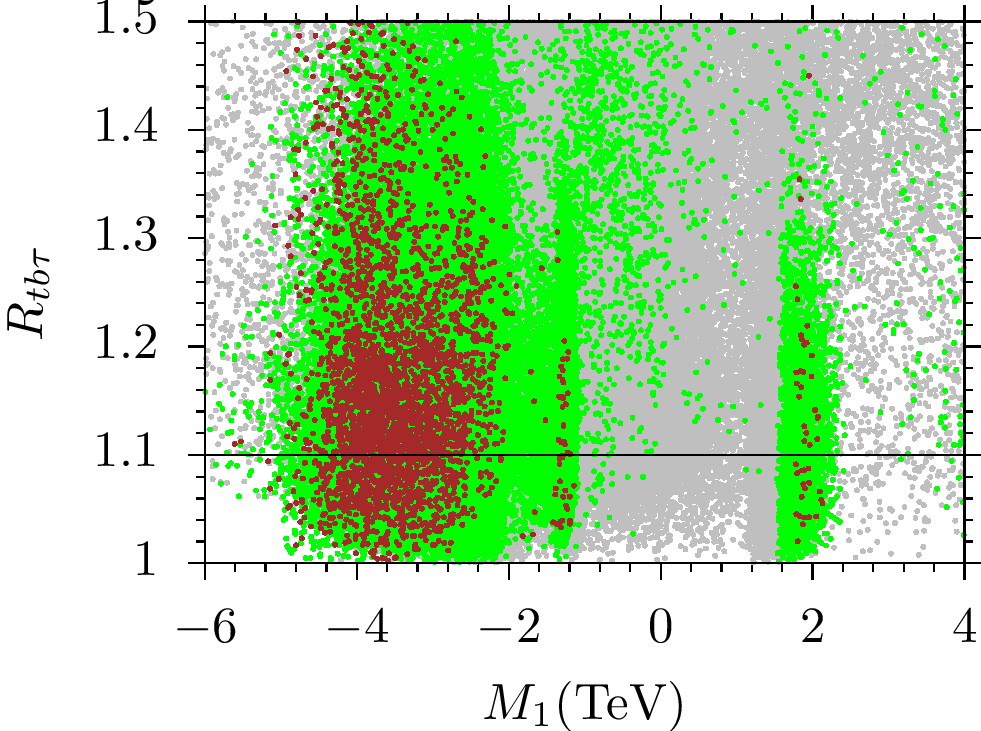}}
\subfigure{\includegraphics[scale=1]{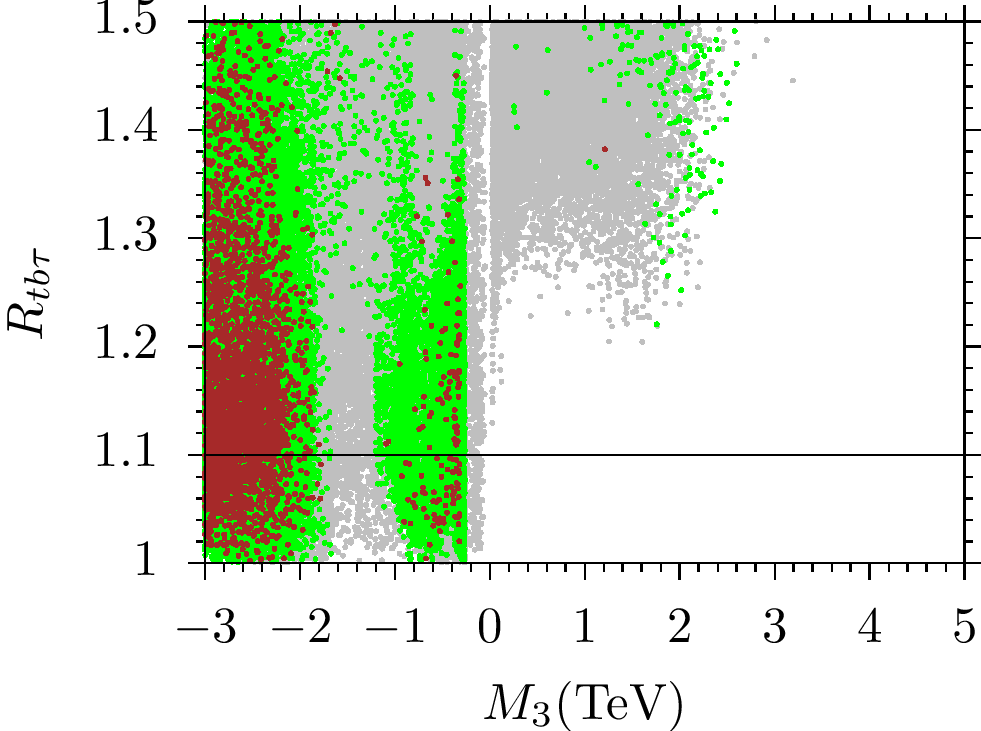}}
\caption{Plots in the $R_{tb\tau}-x_{LR}$, $R_{tb\tau}-y_{LR}$, $R_{tb\tau}-M_{1}$ and $R_{tb\tau}-M_{3}$ planes. The color coding is the same as in Figure \ref{fig1}.}
\label{fig2}
\end{figure}

\begin{figure}[ht!]
\centering
\subfigure{\includegraphics[scale=1]{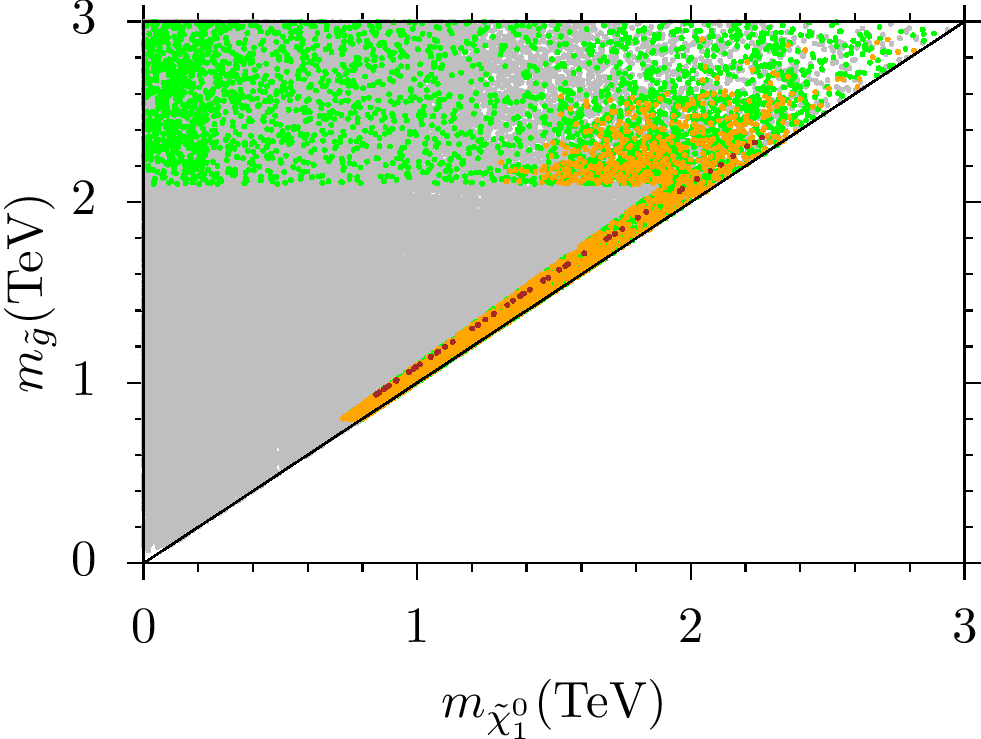}}
\subfigure{\includegraphics[scale=1]{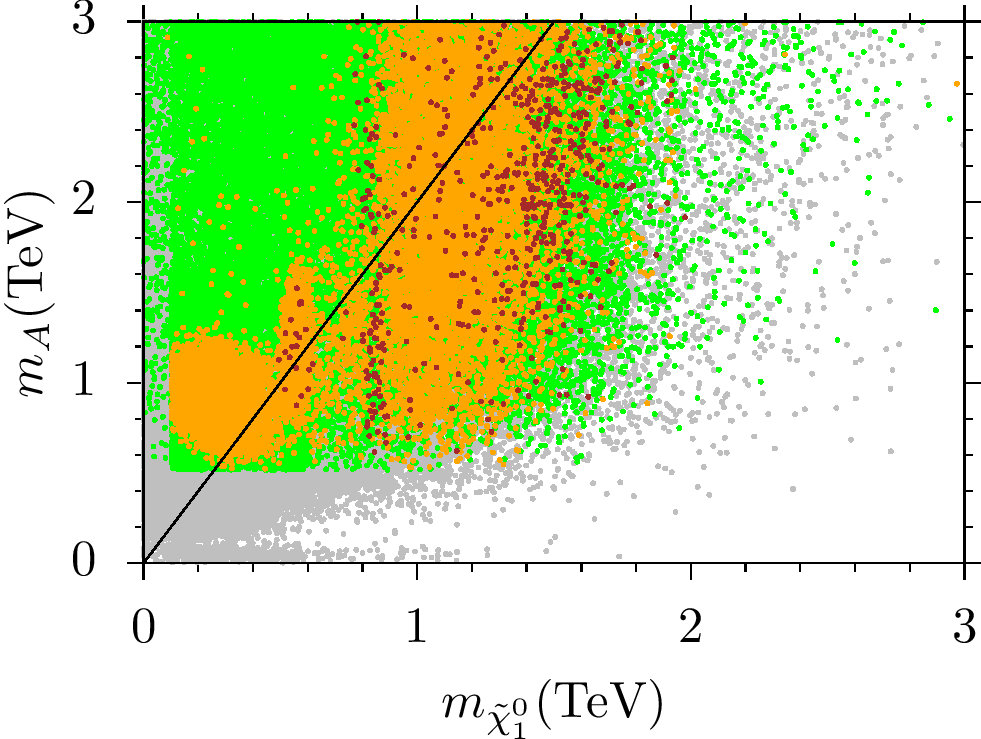}}
\subfigure{\includegraphics[scale=1]{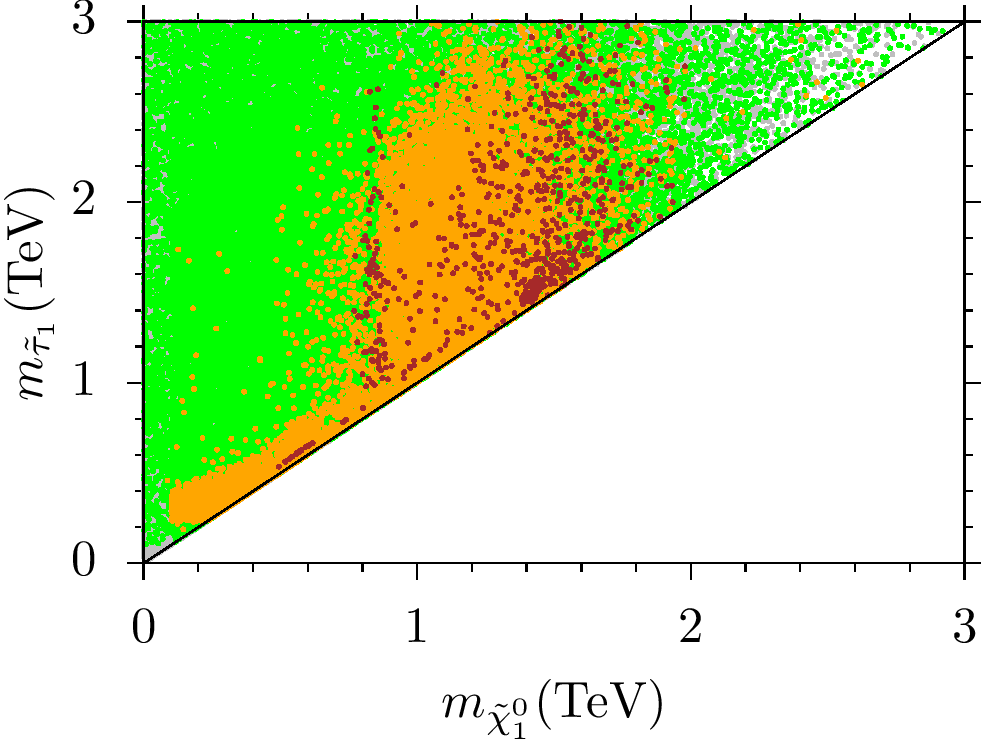}}
\subfigure{\includegraphics[scale=1]{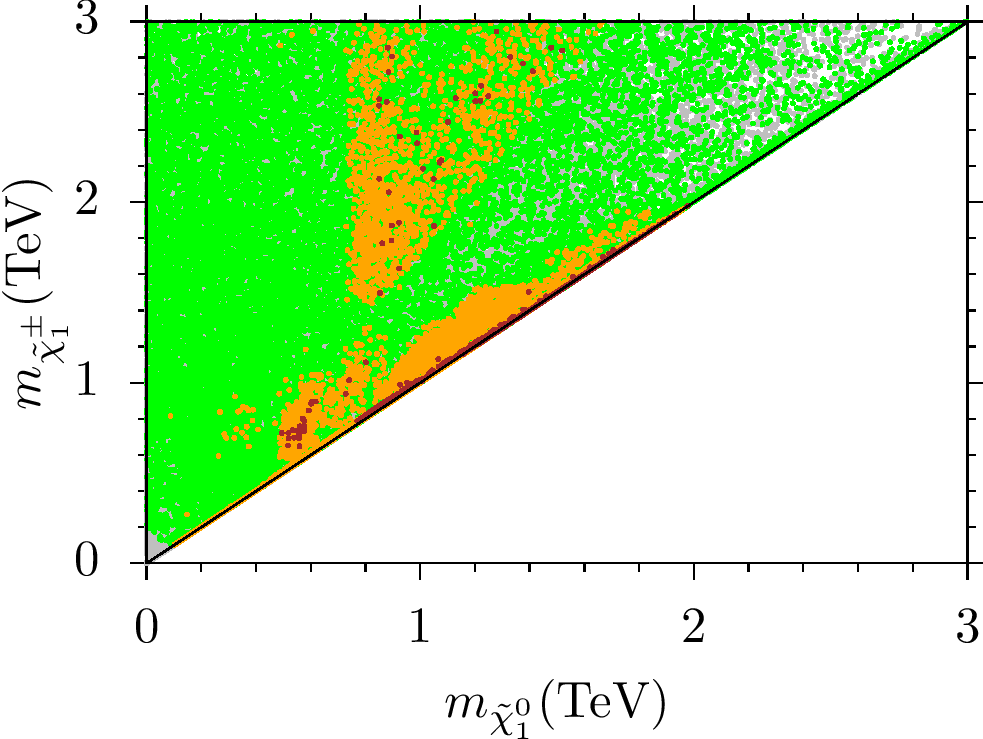}}
\caption{Plots in the $m_{\tilde{g}}-m_{\tilde{\chi}_{1}^{0}}$, $m_{A}-m_{\tilde{\chi}_{1}^{0}}$, $m_{\tilde{\tau}_{1}}-m_{\tilde{\chi}_{1}^{0}}$ and $m_{\tilde{\chi}_{1}^{\pm}}-m_{\tilde{\chi}_{1}^{0}}$ planes. All points are compatible with the REWSB and LSP neutralino conditions. Green points satisfy the mass bounds and constraints from {rare} $B-$meson decays. Orange points form a subset of green and they are compatible {with} $\tbta$ Yukawa unification. Brown points are a subset of orange, they are consistent with the Planck bound on the relic abundance of LSP neutralino within $5\sigma$. The diagonal lines indicate {regions} in which the displayed particles are degenerate in mass, {except for the line} in the $m_{A}-m_{\tilde{\chi}_{1}^{0}}$ plane {which shows} the solutions with $m_{A}=2m_{\tilde{\chi}_{1}^{0}}$.}
\label{fig3}
\end{figure}

\begin{figure}[ht!]
\centering
\subfigure{\includegraphics[scale=1]{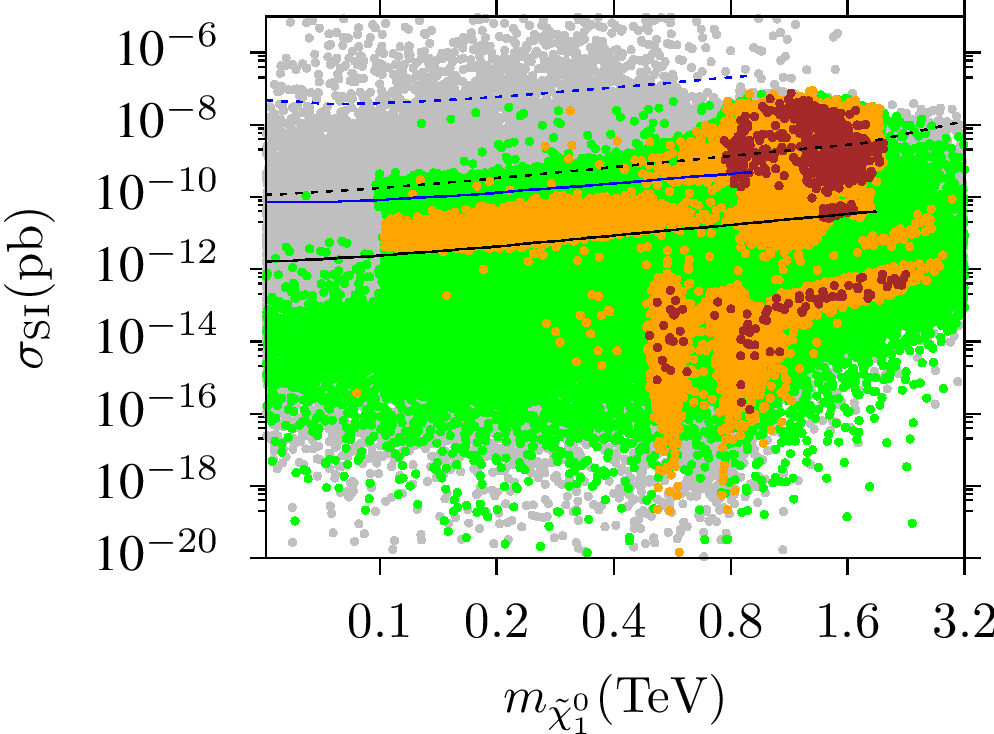}}
\subfigure{\includegraphics[scale=1]{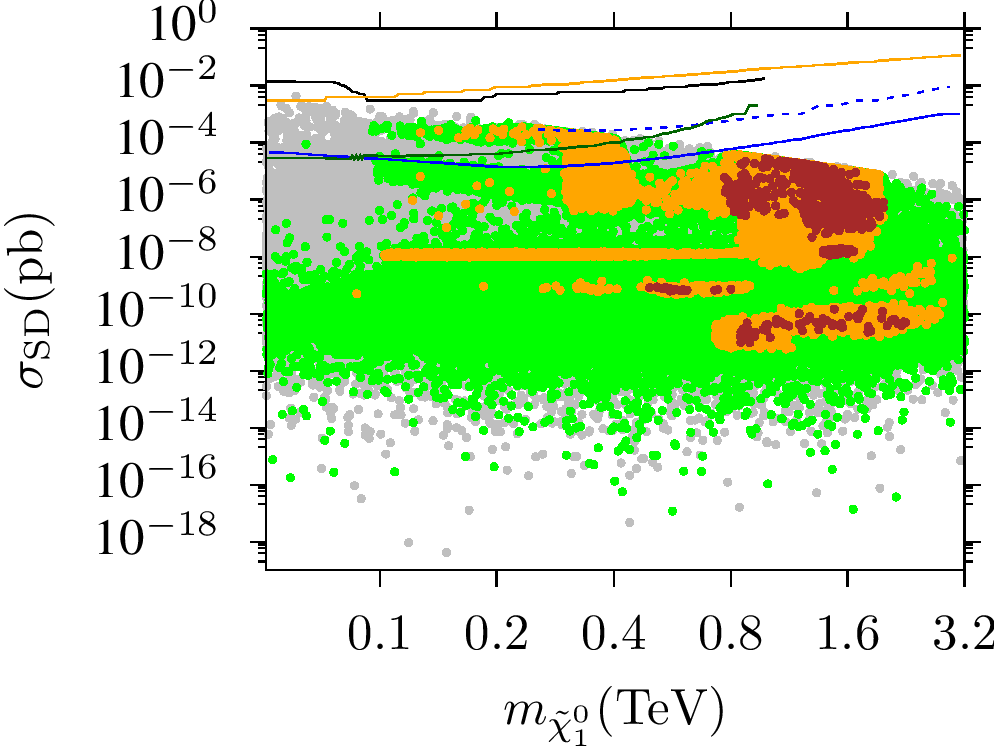}}
\caption{Spin-independent (left) and spin-dependent (right) scattering cross-sections versus the LSP neutralino mass. In the $\sigma_{SI}-m_{\tilde{\chi}_{1}^{0}}$ plane, the dashed (solid) blue line represents the current (future) results from the SuperCDMS experiment \cite{Brink:2005ej}. The dashed (solid) black line indicates the current (future) results from the LUX-Zeplin experiment \cite{Akerib:2018lyp}. In the $\sigma_{SD}-m_{\tilde{\chi}_{1}^{0}}$ plane, the solid black line represents the currrent bound from Super-K \cite{Tanaka:2011uf}, {and} the solid orange line is set by the LUX results \cite{Akerib:2016lao}. The green line is obtained {from} collider analyses \cite{Khachatryan:2014rra}, {and  the} dashed (solid) blue line shows the current (future) results from IceCube DeepCore.}
\label{fig4}
\end{figure}

In this section we present our results {for} $\tbta$ YU {within} the \422 framework. Figure \ref{fig1} displays the fundamental parameter space of $\tbta$ YU in terms of the SSB scalar (top panels) and gaugino (bottom panels) mass terms with plots in the $R_{tb\tau}-m_{\tilde{L}}$, $R_{tb\tau}-m_{\tilde{R}}$, $R_{tb\tau}-M_{2L}$ and $R_{tb\tau}-M_{2R}$ planes. All points are compatible {with} REWSB and LSP neutralino conditions. Green points satisfy the mass bounds and constraints {from rare} $B-$meson decays. Brown points form a subset of green and they yield relic abundance of LSP neutralino consistent with the Planck measurements within $5\sigma$. The horizontal line indicates the region with $R_{tb\tau}=1.1$, and points below this line are considered {as being in very good agreement} with the $\tbta$ Yukawa unification. The $R_{tb\tau}-m_{\tilde{L}}$ plane shows that $\tbta$ YU can be realized within a wide range of $m_{\tilde{L}}$ from about 400 GeV to 10 TeV. Two regions can be identified, which are separated {by} a gray area. The gray region between them is excluded mostly by the gluino mass bound, which will be shown in {more detail} later. The SSB mass term for the right-handed scalar particles can be as light as about 1 TeV, {while} $\tbta$ YU condition restricts it at about 15 TeV from above, as seen from the $R_{tb\tau}-m_{\tilde{R}}$ plane. The left-bottom plane displays the SSB mass term for $M_{2L}$, and as is seen, the $\tbta$ YU solutions can be realized in {a wide range of values for} $M_{2L}$ in our scan. {It} is bounded at about 800 GeV from below by the DM constraints. The mass term for the $SU(2)_{R}$ gaugino could lie in a wider range from about $-10$ TeV to 4 TeV consistent with all the LHC and DM constraints. 

The parameters {quantifying} non-universality in the scalar and gaugino {sectors} are shown in Figure \ref{fig2} along with the SSB gaugino mass terms with plots in the $R_{tb\tau}-x_{LR}$, $R_{tb\tau}-y_{LR}$, $R_{tb\tau}-M_{1}$ and $R_{tb\tau}-M_{3}$ planes. The color coding is the same as in Figure \ref{fig1}. Despite its wider range in our scan, the $R_{tb\tau}-x_{LR}$ plane shows {that LR} breaking in the scalar sector can be realized consistent with the DM constraints if $0.8 \lesssim x_{LR}\lesssim 1.6$. Note that $x_{LR}=1$ restores the symmetry in the scalar sector. {On the other hand, the LR breaking in the gaugino sector can be crucial for $\tbta$ YU}, as seen from the $R_{tb\tau}-y_{LR}$ plane {with} $|y_{LR}|$ as large as $3$. Even though $\tbta$ YU {mostly} prefers $y_{LR}$ to be negative, it is {possible to realize positive $y_{LR}$ values} in a relatively small portion of the parameter space, which also leads to negative $M_{2R}$ in most of the parameter space. Its impact can be seen from the $R_{tb\tau}-M_{1}$ plane {where} $M_{1}$ is mostly negative and its magnitude can be as high as about 5 TeV, {while} it is restricted {to} $M_{1}\sim 2$ TeV in the positive region. Since $M_{1}$ controls the {bino} mass at the low scale, such large values of $M_{1}$ prevent bino {from being the} LSP. {The parameter} $M_{3}$ is {shown} in the right-bottom panel, and it is seen {that $\tbta$ YU condition allows only negative $M_{3}$ values}. In this context, according to Eq.(\ref{eq:gauginomasses}), one can expect very large $M_{2L}$ {for} $M_{1} > 0$; thus this region is most likely to realize a {bino} or {Higgsino} DM. 

We present the low scale mass spectrum in Figure \ref{fig3} with plots in the $m_{\tilde{g}}-m_{\tilde{\chi}_{1}^{0}}$, $m_{A}-m_{\tilde{\chi}_{1}^{0}}$, $m_{\tilde{\tau}_{1}}-m_{\tilde{\chi}_{1}^{0}}$ and $m_{\tilde{\chi}_{1}^{\pm}}-m_{\tilde{\chi}_{1}^{0}}$ planes. All points are compatible {with} REWSB and LSP neutralino conditions. Green points satisfy the mass bounds and constraints from {rare} $B-$meson decays. Orange points form a subset of green and they are compatible {with} $\tbta$ YU. Brown points are a subset of orange {and} they are consistent with the Planck bound on the relic abundance of LSP neutralino within $5\sigma$. The diagonal lines {indicate} regions in which the displayed particles are degenerate in mass, {and the line} in the $m_{A}-m_{\tilde{\chi}_{1}^{0}}$ plane {depicts} solutions with $m_{A}=2m_{\tilde{\chi}_{1}^{0}}$. {Some of the} most interesting results {correspond to} NLSP gluino solutions. {In previous studies NLSP gluino masses of order 1 TeV are found}, compatible with $\tbta$ YU {in the presence of LR symmetry}. However, the $m_{\tilde{g}}-m_{\tilde{\chi}_{1}^{0}}$ plane shows that {in our case} one can realize NLSP gluino solutions compatible with $\tbta$ YU {for} gluino mass scales up to about {2.5} TeV. Moreover, as seen from the other {panels} of Figure \ref{fig3}, the mass spectra also favor the $A-$resonance {solution} if $1\lesssim m_{A}\lesssim 3$ TeV, and stau-neutralino and chargino-neutralino coannihilation {solutions} if $0.6\lesssim m_{\tilde{\tau}_{1}}\lesssim 1.6$ TeV. The $m_{\tilde{\chi}_{1}^{\pm}}-m_{\tilde{\chi}_{1}^{0}}$ plane also shows that the lightest chargino is {almost} as light as the LSP neutralino in most of the parameter space. Approximate mass degeneracy between the lightest chargino and LSP neutralino is one of the characteristics {features} of DM {composed of wino or Higgsino}.

{With a wino and/or Higgsino DM}, one can expect large cross-section in the DM scattering processes. {For wino DM}, the scattering {off} nuclei occurs through $SU(2)$ interactions, while Yukawa interactions take part {if} the Higgsino happens to be the LSP. Figure \ref{fig4} shows results for the spin-independent (left) and spin-dependent (right) scattering cross-sections versus the LSP neutralino mass. In the $\sigma_{SI}-m_{\tilde{\chi}_{1}^{0}}$ plane, the dashed (solid) blue line represents the current (future) results from the SuperCDMS experiment \cite{Brink:2005ej}. The dashed (solid) black line indicates the current (future) results from the LUX-Zeplin experiment \cite{Akerib:2018lyp}. In the $\sigma_{SD}-m_{\tilde{\chi}_{1}^{0}}$ plane, the solid black line represents the currrent bound from Super-K \cite{Tanaka:2011uf}, {and} the solid orange line is set by the LUX results \cite{Akerib:2016lao}. The green line is obtained {from} collider analyses \cite{Khachatryan:2014rra}, {and the} dashed (solid) blue line shows the current (future) results from {IceCube/DeepCore}. The region in the $\sigma_{{\rm SI}}-m_{\tilde{\chi}_{1}^{0}}$ plane, which is cut by the dashed black exclusion curve of the current results from the LUX measurements, implies {a} Higgsino DM. Even though {many} solutions are already excluded due to large scattering cross-sections, it is still possible to realize {a} Higgsino DM slightly below this curve. These solutions are expected to be probed in the direct {detection DM} experiments {in the near future}. The solutions between the current and future exclusion curves also yield a {wino} DM, and they are {within reach of} future experiments. {The remaining} solutions {lying} below all the curves represent the DM {in which bino} is involved.

Most of the models  allowed by the constraints listed in Eq.(\ref{constraints}) and 
compatible with $\tbta$ YU  present topologies that cannot be detected at the LHC according to the Smodels analysis. Note that this is the case {for} the subset of {models} satisfying the DM constraint. Although many models predict relatively low SUSY masses, their signals {escape} the LHC bounds as can be understood  from our previous discussions. For instance,  although the constraints on gluino masses are {quite} severe, we can find cases with {gluino} masses of order 800 GeV compatible with $\tbta$ YU. However, the DM {constraint} bound it at about 900 GeV from below, as seen in Figure~\ref{fig3}.  In this region, the gluino happens to be {the} NLSP, and it can decay into a LSP neutralino along with a quark-antiquark pair from the first two families. The bound from these processes is set as $m_{\tilde{g}}\gtrsim 800$ GeV \cite{Aaboud:2017vwy}. Figure \ref{fig3} also shows that the stau can be as light as {500 GeV or so},  consistent with all the constraints and $\tbta$ YU, where the LSP neutralino is formed mostly {from} bino and/or wino. {If the} charginos and second lightest neutralino are allowed to decay into staus, YU with $m_{\tilde{\tau}}\lesssim 350$ GeV and/or $m_{\tilde{\chi}_{1}^{\pm}}\lesssim 1.1$ TeV is  excluded \cite{CMS:2017fdz}. On the other hand, the lightest stau in {MSSM} is mostly {composed of} the right-handed stau, which forbids the {lightest} chargino to decay into a stau. Besides, as seen from the chargino and LSP neutralino masses shown in the right bottom panel of Figure \ref{fig3}, the chargino decay into a LSP neutralino along with a $W-$boson is not kinematically allowed, which also {loosens} the constraints on the chargino and stau masses.

\begin{table}[h!]
\centering
\begin{tabular}{|c|ccccc|}
\hline

& Point 1 & Point 2 & Point 3 & Point 4 & Point 5 \\
\hline
$m_{\tilde{L}}$   & 8031& 9461  & 781 &  1202  & 3714  \\
$M_{1}$   & 1786 & -4810 &-1288 & -3653  & -4502  \\
$M_{2L}$   & 2859 & 3915 &758.5 & 1559  & 2348  \\
$M_{3}$  & -316.7 & -926 &-2904 &  -2802  & -2537  \\
$A_{0}/m_{\tilde{L}}$  & -1.06 & -0.13 &1.57 &  1.16  & 1.30  \\
$\tan\beta$   & 48.0 & 47.8 & 43.8 & 42.6  & 52.4  \\
\hline 
$x_{LR}$   & 1.43 & 0.78 & 0.83 &  1.45  & 0.84  \\
$y_{LR}$   & 1.11 & -1.89 & -0.28 & -2.71  & -2.47  \\
$m_{\tilde{R}}$   & 11460 & 7398 & 646.9 & 1740  & 3111 \\
$M_{2R}$   & 3188 & -7399 & -210.1 &  -4221  & -5811  \\
\hline
$\mu$   & 8416 & 6401 & 3431 &  3398  & {\color{red} 894.4}  \\ 
\hline
$m_{h}$  & 124.3 & 123.2 & 123.2 & 123.1  & 124.1  \\
$m_{H}$   & 6205 & 5568 & 1088 & 1549   &  2432  \\
$m_{A}$   & 6164 & 5531 & {\color{red} 1081} & 1539 & 2417  \\
$m_{H^{\pm}}$   & 6206 & 5568 & 1093 & 1552  & 2434  \\
\hline
$m_{\tilde{\chi}_{1}^{0}}, m_{\tilde{\chi}_{2}^{0}}$   & {\color{red} 849.6}, {2530} & {\color{red}2259}, 3438 & {\color{red} 558.1}, 711.6 & {\color{red} 1387}, 1638  & {\color{red} 850.6}, {\color{red} 853.3}  \\
$m_{\tilde{\chi}_{3}^{0}}, m_{\tilde{\chi}_{4}^{0}}$   & 7757, 7757 & 5962, 5962 & 3190, 3190  & 3165, 3165  & 2044, 2058  \\
$m_{\tilde{\chi}_{1}^{\pm}}, m_{\tilde{\chi}_{2}^{\pm}}$   & 2535, 7710 & 3439, 5962 & 713.8, 3162 & {\color{red} 1389}, 3137  & {\color{red} 871.5}, 2018  \\
\hline
$M_{\tilde{g}}$   & {\color{red} 933.9} & {\color{red} 2357} &5954 & 5781  & 5360  \\
$m_{\tilde{u}_{L}}, m_{\tilde{u}_{R}}$   & 8183, 11379 & 9906, 7462 & 5172, 5145 &  5163, 5294 & 6019, 5540 \\
$m_{\tilde{t}_{1}}, m_{\tilde{t}_{2}}$   & 4075, 9113 & 4763, 7707 &4371, 4506 &  4331, 4564  & 3523, 4338  \\
\hline
$m_{\tilde{d}_{L}}, m_{\tilde{d}_{R}}$   & 8183, 11558 & 9906, 7623 & 5173, 5151 & 5164, 5258  &  6019, 5516 \\
$m_{\tilde{b}_{1}}, m_{\tilde{b}_{2}}$   & 4131, 8784 & 3491, 7690 & 4340, 4490 &  4327, 4466  & 3556, 4327  \\
\hline
$m_{\tilde{\nu}_{e,\mu}}, m_{\tilde{\nu}_{\tau}}$   & 8104, 6498 & 9720, 8794 & 909.3, 905.6  & {1668}, 1449  & 4042, 3245  \\
$m_{\tilde{e}_{L}}, m_{\tilde{e}_{R}}$   & 8110, 11641 & 9714, 7811 & 914.7, 843.7  & 1667, 2236  & 4043, 3600 \\
$m_{\tilde{\tau}_{1}}, m_{\tilde{\tau}_{2}}$   & 6509, 9396 & 5167, 8783 & 605.6, 988.1  & {\color{red}1436}, 1896  & 1178, 3245  \\
\hline
$\sigma_{{\rm SI}}~(pb)$   & $0.40\times 10^{-14}$ & $0.71\times 10^{-12}$ & $0.13\times 10^{-13}$  & $0.47\times 10^{-10}$  & $0.19\times 10^{-9}$  \\
$\sigma_{{\rm SD}}~(pb)$   & $0.16\times 10^{-10}$ & $0.48\times 10^{-10}$ & $0.68\times 10^{-9}$  & $0.17\times 10^{-7}$  & $0.29\times 10^{-6}$ \\
$\Omega h^{2}$   & 0.116 & 0.124 & 0.122  & 0.120  & 0.125 \\
\hline
$R_{tb\tau}$   & 1.04 & 1.08 & 1.09  & 1.08   & 1.09  \\
\hline
\end{tabular}
\caption{{Benchmark points are compatible with all experimental constraints used in this paper}. All points are chosen to be allowed by the constraints. All masses are given in GeV. {Points 1 and 2} depict NLSP gluino solutions {and point} 3 represents an $A-$resonance solution. The first three points predict  bino-like DM. Point 4 displays {a stau-neutralino coannihilation solution with a wino-like DM}. Point 5 is {a solution with} a Higgsino-like DM {and} with $m_{\tilde{\chi}_{1}^{\pm}}\sim m_{\tilde{\chi}_{2}^{0}}\sim m_{\tilde{\chi}_{1}^{0}}$.}
\label{table1}
\end{table}

Before concluding the discussion about $\tbta$ YU, we present {five} benchmark points {that are compatible with experimental constraints}. All masses are given in GeV. Point 1 depicts a solution for the gluino-neutralino coannihilation scenario with {NLSP} gluino mass {of} about 2.4 TeV. {In contrast, point 2 displays a relatively light NLSP gluino}. Point 3 represents an $A-$resonance solution. The DM relic abundance is saturated with {bino}-like LSP neutralino in the first three points. Point 4 displays a wino-like DM solution where the lightest chargino mass is close to the wino and also to the lightest stau mass. Coannihilations of three species lead to a correct relic DM abundance. Point 5 is a typical solution {for} Higgsino-like DM {with} the chargino and neutralino masses {given} as $m_{\tilde{\chi}_{1}^{\pm}}\sim m_{\tilde{\chi}_{2}^{0}}\sim m_{\tilde{\chi}_{1}^{0}}$. {The spin-independent} cross-section is calculated to be slightly below the current LUX result and {so this scenario} will be tested relatively soon.

\section{Conclusion}
\label{sec:conc}

We have explored the LHC and DM implications of $\tbta$ YU in the supersymmetric \422 framework without imposing a discrete LR symmetry. We only accept solutions which yield one of the neutralinos as the LSP {that saturates the DM abundance. We identify} the gluino-neutralino coannihilation scenarios, and {present consistent solutions for} $m_{\tilde{g}}\lesssim 2.4$ TeV. {Without the NLSP constraint the gluino can} be as heavy as about 6 TeV, which can be probed {at the} LHC {and future colliders}. In addition to the gluino-neutralino coannihilation {scenario, some} $A-$resonance solutions are also identified {with} $1 \lesssim m_{A} \lesssim 3$ TeV. Similarly, the stau-neutralino and chargino-neutralino coannihilation processes can be realized {with} the stau and chargino {masses in the range} $0.6 \lesssim m_{\tilde{\tau}}, m_{\tilde{\chi}_{1}^{\pm}}\lesssim 2$ TeV. 

{The} \422 {model also yields wino} and Higgsino-like DM {as well as solutions with bino DM}. We observe that {while many} of the Higgsino DM solutions {are} excluded by the direct {detection} experiments, {it} is still possible to realize {some} solutions {lying slightly} below the current exclusion curves. In {other words}, Higgsino DM in the \422 framework {will be seriously} tested {in the near future}. Wino-like DM solutions are allowed by the current measurements, and they {lie within} the reach of the near future experiments. 

We exemplify our findings with five benchmark points including the {full} spectrum for the SUSY particles. In addition to the DM implications, the stop cannot be lighter than about 3 TeV, while squarks of the first two families {are relatively heavy} ($\gtrsim 5$ TeV).

\section*{Acknowledgments}
We would like to thank Zafer Alt\i n for discussions. CSU also thank Universidad de Huelva and Bartol Research Institute of University of Delaware for their kind hospitality, where part of this work has been done. The research of M.E.G. was supported by the Spanish MINECO, under grant  FPA2017-86380-P. R. Q.S. acknowledges support by the DOE grant No. DE-SC0013880. The work of CSU is supported in part by the Scientific and Technological Research Council of Turkey (TUBITAK) Grant no. MFAG-118F090. Part of the calculations reported in  this paper were performed at the National Academic Network and Information Center (ULAKBIM) of TUBITAK, High Performance and Grid Computing Center (TRUBA Resources). CSU also acknowledges support from the CEAFMC of the University of Huelva.

\end{document}